\documentclass[prd,preprint,superscriptaddress,amsmath,amssymb]{revtex4}

\usepackage{slashed,bbold}
\usepackage{soul}

\usepackage{graphicx,color}

\begin{document}

{\begin{flushright}{KIAS-17054}
\end{flushright}}

\title{ Penguin $b\to s\ell'^+ \ell'^-$ and $B$-meson anomalies  in a gauged ${L_\mu -L_\tau}$}
%
%
\author{Chuan-Hung Chen}
\email{physchen@mail.ncku.edu.tw}
\affiliation{Department of Physics, National Cheng-Kung University, Tainan 70101, Taiwan}

\author{Takaaki Nomura}
\email{nomura@kias.re.kr}
\affiliation{School of Physics, KIAS, Seoul 130-722, Korea}

\date{\today}

\begin{abstract}
The $Z'$-gauge boson in an $U(1)_{L_\mu-L_\tau}$ gauge symmetry  has  two interesting features: one is its vector couplings to the charged leptons, and the other is the decoupling from the electron. Based on these properties, we investigate the feasibility to simultaneously resolve the $R_{K^{(*)}} = BR(B\to K^{(*)} \mu^+ \mu^-)/BR(B\to K^{(*)} e^+ e^-)$ and $R_{D^{(*)}} = BR(\bar B\to D^{(*)} \tau \bar\nu_\tau)/BR(\bar B\to D^{(*)} \ell \bar\nu_\ell)$ anomalies in an $U(1)_{L_\mu-L_\tau}$ model, where the former  is expected  to arise from  the $Z'$-penguin-induced $b\to s \mu^+ \mu^-$ process and the latter  from the tree-level $b\to c \tau \bar\nu_\tau$ decay. In order to achieve the intended purpose, we employ one vector-like doublet lepton and  one singlet scalar leptoquark (LQ), in which the new particles all carry the $U(1)_{L_\mu-L_\tau}$ charges;  the $b\to sZ'$ effective interaction is generated from the  vector-like lepton and LQ loop, and the $b\to c \tau \bar\nu_\tau$ decay is induced from the LQ. When the constraints from the  $b\to s \gamma$, $B^+\to K^+ \nu \bar\nu$, $B^-_c\to \tau \bar \nu_\tau$,  $\Delta F=2$, and $\tau\to \mu \ell \bar \ell$ processes are included, it is found that $R_D$ and $R_{D^*}$ can be enhanced to fit the experimental data, and the Wilson coefficient $C_9$ from the LQ-loop can reach $C^{LQ,\mu}_9\sim -1$, which can explain the $R_{K}$ and $R_{K^*}$ anomalies. In addition, in this simple model, the Higgs lepton-flavor violating $h\to \mu \tau$ decay can occur at the tree level, and its branching ratio can be as large as the current experimental upper limit.

\end{abstract}
\maketitle

\section{Introduction}

A $2.6\sigma$ deviation from the standard model (SM) prediction in $R_{K} = BR(B^+\to K^+ \mu^+ \mu^-)/BR(B^+\to K^+ e^+ e^-)$ with $R_{K}=0.745^{+0.090}_{-0.074} \pm 0.036$  was earlier reported by the LHCb collaboration in~\cite{Aaij:2014ora}, where $BR(B^+\to K^+ \ell^+ \ell^-)$ ($\ell=e,\mu$) denotes the branching ratio (BR) of the $B$ decay; the integrated dilepton invariant mass square range is $1<q^2<6$ GeV$^2$, and the SM prediction is unity in that region~\cite{Hiller:2003js}.  Intriguingly,  LHCb recently finds a similar deviation  in $R_{K^*} = BR(B^0\to K^{*0} \mu^+ \mu^-)/BR(B^0\to K^{*0} e^+ e^-)$ as~\cite{Aaij:2017vbb}:
 \begin{align}
 R_{K^*} =\left\{
\begin{array}{c}
  0.66^{+0.11}_{-0.07} \pm 0.03 \  \ \ {\rm for} \ 0.0045< q^2 < 1.1 \ {\rm GeV^2}\,,   \\
 0.69^{+0.11}_{-0.07} \pm 0.05 \  \ \ {\rm for} \ \  \ \  \ 1.1< q^2 < 6.0 \ {\rm GeV^2}\,. \end{array}
\right.
 \end{align}
 The SM prediction at leading order is $R_K \approx R_{K^*} \approx 1$. When  QED radiative corrections are included, it is found that the influence on $R_{K,K^*}$ does not exceed a few percent~\cite{Bordone:2016gaq}. Thus, the importance of $R^{\rm exp}_{K^{(*)}} < R^{SM}_{K^{(*)}}$ indicates a violation of lepton-flavor universality~\cite{Hiller:2003js}.

 The lepton-flavor universality is also confronting a test in the terms of $b\to c \tau \bar \nu_\tau$ decay.  BaBar~\cite{Lees:2012xj,Lees:2013uzd}, Belle~\cite{Huschle:2015rga,Abdesselam:2016cgx,Hirose:2016wfn}, and LHCb~\cite{Aaij:2015yra,LHCbDv} experimentally  observed excesses in  the ratios of  $BR(\bar B \to D^{(*)} \tau \bar\nu_\tau)$ to $BR(\bar B \to D^{(*)} \ell \bar\nu_\ell)$, and the averaged values  are obtained as~\cite{Amhis:2016xyh}:
\begin{align}
R_{D}  &=  0.407 \pm 0.039 \pm 0.024\,,  \nonumber \\
 R_{D^*}  &=  0.304 \pm 0.013 \pm 0.007\,,
\end{align}
 where  the SM predictions using different approaches are closed to each other, and they are  given as: $R_{D} = 0.299 \pm 0.011$ \cite{Lattice:2015rga}, $ R_{D}=0.300 \pm 0.008$~\cite{Na:2015kha} in lattice calculations, and  $R_{D} = 0.299 \pm 0.003$~\cite{Bernlochner:2017jka}; $R_{D^*} = 0.252 \pm 0.003$~\cite{Fajfer:2012vx}, $R_{D^*}=0.262 \pm 0.010$~\cite{Bigi:2017jbd},  and $R_{D^*} = 0.257 \pm 0.003$~\cite{Bernlochner:2017jka}.  The measurements of $R_D$ and $R_{D^*}$ exceed the SM results by  around $2.3\sigma$ and $3.4\sigma$, respectively. When the correlation between $R_D$ and $R_D^*$ is considered, the deviation from the SM is about $4.1\sigma$~\cite{Amhis:2016xyh}.   Based on these observations, various interesting extensions of the SM have been proposed to explain anomalies~\cite{Altmannshofer:2014cfa,Crivellin:2015mga,Altmannshofer:2015mqa,Altmannshofer:2016jzy,Ko:2017yrd,Chen:2017hir,Chen:2017eby,Megias:2017ove,Altmannshofer:2017fio,Crivellin:2017zlb,Capdevila:2017bsm,Altmannshofer:2017yso,Celis:2017doq,Kamenik:2017tnu,Alok:2017jaf,Alok:2017sui,Bonilla:2017lsq,Tang:2017gkz,Chiang:2017hlj,Choudhury:2017qyt,Crivellin:2017dsk,Cline:2017ihf,Guo:2017gxp,Khalil:2017mvb,Buttazzo:2017ixm,Ko:2017quv,Datta:2017ezo,Altmannshofer:2017poe,Kawamura:2017ecz,Ko:2017lzd,Matsuzaki:2017bpp,Sumensari:2017mud,Alonso:2017uky,Alonso:2017bff,Cheung:2016frv,Alonso:2016oyd,Cheung:2016fjo,Becirevic:2016yqi,Bauer:2015knc,Das:2017kfo,Das:2016vkr,Li:2016vvp,Hiller:2017bzc,Belanger:2015nma,Carmona:2015ena,Chauhan:2017ndd,Boucenna:2016wpr,Boucenna:2016qad,Descotes-Genon:2013wba,Crivellin:2015lwa,Dorsner:2017ufx,Dorsner:2016wpm,Becirevic:2015asa,Dorsner:2013tla,Ciuchini:2017mik,Bardhan:2016uhr,Bardhan:2017xcc,Ghosh:2017ber,Cheung:2017efc,Chen:2016dip,Bian:2017rpg,Akeroyd:2017mhr}.

 It is known that the $\bar B \to D^{(*)} \ell' \bar\nu_{\ell'}$ ($\ell' = e,\mu,\tau$) decays are the $W$-mediated tree processes in the SM; however, the $ B\to K^{(*)} \ell'^+ \ell'^-$ decays are  flavor-changing neutral current (FCNC) processes, and are generated at the one-loop level, including box and penguin diagrams.  If the $R_{D^{(*)}}$ and $R_{K^{(*)}}$ anomalies simultaneously arise from the tree diagrams from the same source (e.g., a scalar leptoquark),  inevitably, they will encounter the strict bounds from the $\Delta F=2$~\cite{Altmannshofer:2014cfa,Bauer:2015knc,Alok:2017sui,Chen:2017hir}, $B^-_c\to \tau \bar \nu_\tau$~\cite{Li:2016vvp,Alonso:2016oyd,Akeroyd:2017mhr}, and $b\to s \nu \bar \nu$ decays~\cite{Bauer:2015knc,Becirevic:2016yqi,Chen:2017hir}.  Therefore, if a unified resolution to $R_{D^{(*)}}$ and $R_{K^{(*)}}$ is from  the tree level, which is the approach most frequently used in the literature, it is better that the interactions involved are from different media. 
 
 In this study,  we propose that, like the situation in the SM,  the $b\to s \mu^+ \mu^- (\tau^+ \tau^-)$ decay arises from a penguin diagram, whereas the $b\to c \tau \bar\nu_\tau$ decay is produced through a tree-level charged current. It is found that the proposed effects can be easily achieved in a gauged $U(1)_{L_\mu -L_\tau}$ model, where the associated $Z'$-gauge boson only couples to the $\mu$- and $\tau$-lepton, but not to the electron~\cite{He:1991qd, He:1990pn}.  In the literature, an $U(1)_{L_\mu - L_\tau}$ model with vector-like quarks (VLQs), which is used to resolve the $B\to K^{(*)} \mu^+ \mu^-$ anomalies arising from tree effects, was studied in~\cite{Altmannshofer:2014cfa,Altmannshofer:2015mqa,Altmannshofer:2016jzy}.  Although the loop-induced $b\to s \mu^+ \mu^-$ decay can be generated by the   $Z_2$-odd VLQs and scalar~\cite{Ko:2017yrd}, the new physics effects cannot be applied to the $b\to c \tau \bar\nu_\tau$ decay. The authors in~\cite{Crivellin:2015mga}  resolved the 
 $b \to s \mu^+ \mu^-$ decay issue with the gauged $L_\mu -L_\tau$ symmetry when doublet and singlet VLQs and a second Higgs doublet  were introduced to the model. In spite of the attractive implications on other flavor physics, the FCNC $b\to s \mu^+ \mu^-$ process was induced at the tree level through the mixing between VLQs and the SM quarks.  Here, we provide an alternative version without the vector-like quarks and second Higgs doublet.

 The remainder of this paper is organized as follows. We introduce the model and the resulting effective interactions for $b\to c \tau \bar \nu$ and $b\to s \ell'^+ \ell'^{-}$ in section II. We discuss possible potential constraints in section III, where they include the neutrino trident production, $b\to s \gamma$, $B_c \to \tau \nu$, $\Delta F=2$, and $\tau\to \mu \ell \bar \ell$ processes.  The physical implications on $R_{D^{(*)}}$, $C_9$ Wilson coefficient, $h\to \mu \tau$ decay, and muon anomalous magnetic dipole moment ( muon $g-2$) are also shown in this section. The summary is given in section IV.  
 
  \section{ The model and the effective interactions for $b\to c \tau \bar \nu_\tau$ and $b\to s \ell'^+ \ell'^-$}

 Since violation of lepton-universality concerns lepton properties, we assume that  the breaking effects only occur in the lepton sector; that is, only leptons or particles carrying a lepton-number can have the  $U(1)_{L_\mu - L_\tau}$ charges. If the flavor-changing $b\to s Z'$ and $b\to c \tau \bar\nu_\tau$ arise from the same origin, the most promising mediator is a leptoquark (LQ), which can couple to a quark and a lepton at the same vertex.  Thus, in order to generate the $b\to s Z'$ and $b\to c \tau \bar\nu_\tau$ decays from the same mediator via a loop and a tree diagram, respectively,  we  introduce the vector-like doublet lepton $L^T_4=(2,-1/2)$, the singlet scalar LQ $\Phi^{-1/3}=(1,-1/3)$, and  the singlet scalar $S=(1,0)$ under $(SU(2)_L, U(1)_Y)$ gauge symmetry, where $S$ is responsible for the spontaneous $U(1)_{L_\mu-L_\tau}$ symmetry breaking, for which  the $U(1)_{L_\mu-L_\tau}$ charges of the relevant particles are given in Table~\ref{tab:CU1}. Particles not shown in the table have no $U(1)_{L_\mu-L_\tau}$ charges.   The earlier studies using the same singlet leptoquark to explain  the $R_{D^{(*)}}$ anomalies can be found in~\cite{Bauer:2015knc, Altmannshofer:2017poe,Chen:2017hir}.   Accordingly, the gauge invariant Yukawa couplings are expressed as:
 \begin{align}
 -{\cal L}_{Y} & =\bar L_\ell  Y^\ell H \ell_R + \left( \bar L_\tau i\tau_2 {\bf g} P_R Q^c_L  +  \bar \tau_R {\bf  w} P_L u^c_R  + \bar L_{4L} {\bf f} i\tau_2 P_R Q^c_L \right) \Phi^{-1/3} \nonumber \\
 & + y_\tau \bar L_{4L} H \tau_R + y'_\mu \bar L_\mu  L_{4R} S + m_L \bar L_{4L} L_{4R}+H.c.\,, \label{eq:LY}
  \end{align}
  where we  suppressed the quark-flavor indices, $L^T_\ell=(\nu_\ell, \ell)_L$ and $Q^T_L=(u,d)_L$ are the left-handed doublet lepton and quark, respectively, $Q^c_L$ is the charge-conjugate of $Q_L$, $L^T_4\equiv (N_{\tau'}, \tau')$, and $H$ is the SM Higgs doublet. 
  From Eq.~(\ref{eq:LY}), it can be seen that only $\tau$ and vector-like leptons can couple to the scalar LQ.  Since the scalar potential was discussed in~\cite{Chen:2017cic,Chen:2017gvf}, we skip this explanation. If we take $S=(v_S + \phi_S)/\sqrt{2}$ with $v_S$ being the vacuum expectation value (VEV) of $S$, the $Z'$ mass can be obtained as $m_{Z'}=2 g_{Z'} v_S$ in this model~\cite{Chen:2017cic}. Although these new Yukawa couplings generally are complex, in the following analysis, we take these parameters to be real numbers.  

\begin{table}[hptb]
\caption{$U(1)_{L_\mu-L_\tau}$ charges of involving leptons, scalar LQ, and $S$.}
\begin{tabular}{ccccccc} \hline \hline
            &  ~~~$e$~~~  & ~~~$\mu$~~~ & ~~~ $\tau$ ~~~& ~~~ $L_4$  ~~~ &~~~$\Phi^{-1/3}$~~~&~~~$S$~~~\\ \hline 
 ${L_\mu -L_\tau}$ &   0   &   1      & $-1$      &  $-1$  & $-1$ &  $2$ \\ \hline  \hline

\end{tabular}
\label{tab:CU1}
\end{table}

\begin{figure}[hpbt]
\includegraphics[width=120mm]{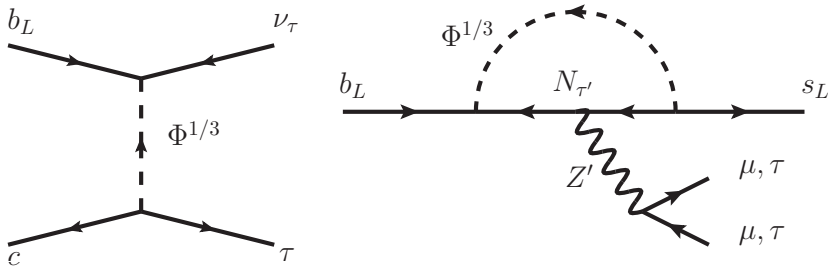}  
\caption{ Sketched Feynman diagrams for the   $b\to c \tau \bar\nu_\tau$ and $b\to s (\mu^+ \mu^-, \tau^+ \tau^-)$ decays.}
\label{fig:one-loop}
\end{figure}

 If we decompose Eq.~(\ref{eq:LY}) and use the quark mass eigenstates, the relevant Yukawa couplings can be written as:
 \begin{align}
 -{\cal L}_{Y} &  \supset \left( \bar \nu_\tau {\bf g} P_R d^c_L  - \bar \tau {\bf  g'} P_R u^c_L  + \bar \tau {\bf w} P_L u^c_R \right) \Phi^{-1/3} + \left( \bar N_{\tau'} {\bf f} P_R d^c_L  - \bar \tau' {\bf  f'} P_R u^c_L   \right) \Phi^{-1/3} \nonumber \\
 & + \frac{y_\tau}{\sqrt{2}}(v + h) \bar \tau' P_R \tau + \frac{y'_\mu}{\sqrt{2}} (v_S + \phi_S) \left( \bar \nu_\mu P_R N_{\tau'} + \bar \mu P_R \tau' \right) + H.c.\,, 
 \label{eq:Yukawa}
 \end{align}
 where  ${\bf g}' \equiv \bf {gV}^T$, ${\bf f}' \equiv {\bf fV}^T$,  ${\bf V}= U^{u}_{L} U^{d\dagger}_L$ denotes the Cabibbo-Kobayashi-Maskawa (CKM) matrix, $U^{u,d}_{L}$ are the quark-flavor mixing matrices for diagonalizing the quark mass matrices, and the flavor mixing matrices of $U^d_L$ and $U^u_R$ can be absorbed into ${\bf g}({\bf f})$ and ${\bf w}$, respectively.  Following Eq.~(\ref{eq:Yukawa}), the tree diagram for $b\to c \tau \bar\nu_\tau$ and the penguin diagram for $b\to s \mu^+ \mu^-(\tau^+ \tau^-)$  are sketched in Fig.~\ref{fig:one-loop}. Accordingly, the effective Hamiltonian for $b\to c \tau \bar\nu_\tau$  can be written as:
\begin{align}
 {\cal H}^{LQ}_{b\to c} &=  \frac{g_2 g'^*_3 }{2m^2_{\Phi}} \bar c \gamma_\mu P_L b \bar \tau \gamma^\mu P_L \nu_\tau \nonumber \\
 & + \frac{ w_2 g^*_{3} }{2m^2_{\Phi}}  \left(-  \bar c P_L b \, \bar\tau P_L \nu_{\tau} + \frac{1}{4} \bar c \sigma_{\mu\nu} P_L b \, \bar \tau \sigma^{\mu \nu} P_L \nu_{\tau} \right) \,, \label{eq:Hbc} 
\end{align}
where $m_\Phi$ is the LQ mass, and the Fierz transformations have been used.  The four-Fermi interactions for the penguin $b\to s \ell'^+ \ell'^-$ can be formulated as:
 \begin{align}
 {\cal H}^{LQ} _{b\to s}& = - \frac{G_F V^*_{ts} V_{tb}}{\sqrt{2}} \frac{\alpha_{\rm em}}{\pi} C^{LQ,\ell'}_9 \bar s \gamma_\mu P_L b \bar \ell' \gamma^\mu \ell' \,,  \label{eq:HbsLQ}\\
 C^{LQ,\ell'}_9 & = \frac{f_2 f^*_3 X_{\ell'}}{(4\pi)^2  C_{SM}} \frac{g^2_{Z'} m^2_L}{m^2_\Phi(q^2 - m^2_{Z'} )} J_0\left( \frac{m^2_L}{m^2_\Phi}\right)  \,,  \label{eq:CZp}\\
 J_0(x) & =  \frac{1}{x-1} - \frac{\ln x}{(x-1)^2}\,, \nonumber
  \end{align}
where $X_{e,\mu,\tau}=(0, 1,-1)$ are the lepton $U(1)_{L_\mu-L_\tau}$ charges,  $C_{ SM}=G_F V^*_{ts} V_{tb} \alpha_{\rm em}/(\sqrt{2} \pi)\approx -8.14 \times 10^{-10}$ GeV$^{-2}$,  and  $q^2$ is the dilepton invariant mass.  It can be seen that like the enhancement  factor $m^2_t/m^2_W$ in the SM, we have the potential enhancement factor $m^2_L/m^2_{\Phi}$ in $C^{LQ,\ell'}_9$. Although the $Z'$-gauge boson can emit from the LQ inside the loop, since the diagram is suppressed by $m_b/m_L$, we have ignored its contribution. We note that  the lepton current in Eq.~(\ref{eq:HbsLQ}) has no axial-vector current, the $B_s\to \mu^+ \mu^-$ decay cannot provide a strict bound on the parameters. 
 In the following discussions, we focus on $C_9^{LQ, \mu}$  in Eq.~(\ref{eq:CZp}) since we consider the $b \to s \mu^+ \mu^-$ process.

\section{Phenomenological discussions and numerical analysis}

In the following numerical estimations, we take the values of the  parameters as:
 \begin{align}
 & G_F \approx 1.166\times 10^{-5}\, {\rm GeV}^{-2}\,, ~ V_{cb}\approx 0.04\,, ~ V_{ts}\approx -0.04\,, ~V_{tb} \approx 1\,,  \nonumber \\
 & m_t \approx 174\, {\rm GeV}\,, ~ m_{B,B_s, B_c}\approx ( 5.28, 5.37, 6.28)\, {\rm GeV}\,, ~ m_{b(c)}\approx 4.6 (1.3)\, {\rm GeV}\,.
 \end{align}
 For illustrating the constraints of new free parameters, we fix the LQ mass to be $m_\Phi =1000$ GeV, unless stated otherwise.

\subsection{Constraint from the neutrino trident production} \label{sec:tri}

To calculate the  penguin contribution, we need to know the constraints of  $g_{Z'}$ and $m_{Z'}$. If we focus on $m_{Z'} > 10$ GeV, basically, two main experiments are involved: one is neutrino trident production~\cite{Altmannshofer:2014cfa,Mishra:1991bv}, and the other is the $Z\to 4 \ell$ measurement~\cite{Aad:2014wra,Altmannshofer:2016jzy}. According to the results in~\cite{Altmannshofer:2016jzy}, the limit from the neutrino trident production can be expressed as $g_{Z'}/m_{Z'} < 1.9 \times 10^{-3}$ GeV$^{-1}$ and is stronger than that from the $Z\to 4\ell$ process when $m_{Z'} > 40$ GeV. If we take  $g_{Z'}/m_{Z'}\sim 1.8\times 10^{-3}$ GeV$^{-1}$ and $m_L/m_\Phi\sim 2$, then $C^{Z',\mu}_9 \sim -1.1$, which is used to explain $R_{K^{(*)}}$ anomalies, can be achieved  when $f_{2} f^*_3 \sim 0.06$ is taken. However, we need to further investigate if the required value of $f_{2} f^*_3$ can be satisfied by the current data. 
 We note that when $g_{Z'}/m_{Z'}\sim 10^{-3}$ GeV is used, we have $m_{Z'} \lesssim 3.5$ TeV for $g_{Z'} \leq \sqrt{4 \pi}$ due to the perturbativity requirement. In addition,  due to $v_S = m_{Z'}/(2 g_{Z'})$, we obtain $v_S \geq 263$ GeV  from the neutrino trident production constraint.

\subsection{Constraints from flavor physics and LHC}

 In the following, we discuss the possible constraints from  flavor physics, such as the  $b\to s \gamma$, $b\to s \nu \bar \nu$, $B^-_c\to \tau \bar \nu_\tau$, $\Delta F=2$, and $\tau \to \mu \ell \bar\ell$ processes. Since our motivation is to demonstrate whether $b\to s Z'$ and $b\to c \tau \bar\nu_\tau$, which are induced from the same mediator,  can simultaneously resolve the $R_{K^{(*)}}$ and $R_{D^{(*)}}$ anomalies in the gauged $L_\mu-L_\tau$ model, but not to give global fitting to all experimental data, for simplicity, we will take the irrelevant parameters to be small or non-existent  in the following discussions.

 The similar loop effects for $b\to s Z'$ shown in Fig.~\ref{fig:one-loop} can also contribute to the radiative $b\to s \gamma$ decay, but the photon can only be emitted from the charged LQ.  Hence, the dimension-5 electromagnetic dipole operator from the LQ loop can be easily obtained as:
  \begin{align}
  {\cal H}^{LQ}_{b\to s \gamma} & = - \frac{V^*_{ts} V_{tb}}{\sqrt{2}} C^{LQ}_{7\gamma}  \left[ \frac{m_b e}{4\pi^2} \bar s \sigma_{\mu \nu} P_R b F^{\mu \nu}\right] \,,  \\
  C^{LQ}_{7\gamma} & =-\frac{\sqrt{2}}{G_F V^*_{ts} V_{tb}} \frac{Q_\Phi f_2 f^*_3  }{4 m^2_\Phi} J_1\left( \frac{m^2_L}{m^2_\Phi} \right) \,, \nonumber \\
  J_1(x) &= \frac{1}{12(x-1)^2} + \frac{x(x+2)}{6(x-1)^3} - \frac{x^2 \ln x}{2(x-1)^4}\,, \nonumber
  \end{align}
  where $Q_\Phi=-1/3$ is the electric charge of $\Phi^{-1/3}$.  With the values of parameters used in Sec.~\ref{sec:tri} and  $|f_{2} f^*_{3}|\sim 0.06$ for $C^{LQ,\mu}_9  \sim -1$, 
  we get $|C^{LQ}_{7\gamma}| \sim 4.5\times 10^{-4}$, which is three orders of magnitude smaller than the SM result of $|C^{SM}_{7\gamma}| \sim 0.3$.   Clearly, the  $b\to s \gamma$ decay cannot significantly bound the parameter $f_2 f^*_3 $. 
 
 Next, we discuss the limit from the $b \to s \nu \bar\nu$ decay. Compared with the inclusive decay mode,  the experimental measurement in the exclusive channel is more closer to the theoretical prediction; therefore, we consider the constraint from $B^+\to K^+ \nu \bar \nu$, where the  SM prediction is $BR^{\rm SM}(B^+\to K^+ \nu \bar\nu)\approx 4\times 10^{-6}$~\cite{Buras:2014fpa}, and the current upper limit is $BR^{\rm exp}(B^+\to K^+ \nu \bar\nu)< 1.6 \times 10^{-5}$~\cite{PDG}; that is, the new physics effects can at most enhance the SM result by a factor of 4. In this model, it is found that the $b\to s \nu \bar\nu$ can be induced from tree and loop diagrams. Except where only the left-handed neutrinos are involved, the loop contribution is the same as that for $b\to s \ell'^+ \ell'^-$. Thus, from Eqs.~(\ref{eq:Yukawa}) and (\ref{eq:HbsLQ}),  the tree and loop effective interactions can be combined as:
  \begin{align}
  {\cal H}^{LQ}_{b\to s \nu \bar \nu} & = - C^\nu_{LQ}  (\bar s b)_{V-A} (\bar \nu \nu)_{V-A}\,, \\
  C^{\nu_{\ell'}}_{LQ} & = \frac{g_2 g^*_3}{8 m^2_\Phi}  C_\nu+  \frac{ C^{LQ,\nu_{\ell'}}_9 C_{SM} }{4} \nonumber 
  \end{align}
 with $C_{\nu_e, \nu_\mu, \nu_\tau}=(0,0,1)$, $(\bar f f')_{V-A}=\bar f \gamma_\mu (1-\gamma_5) f'$.
 Since the induced four-Fermi operators are the same as those in the SM, the BR for the $B^+ \to K^+ \nu \bar \nu$ decay can be simply formulated as:
 \begin{align}
  BR(B^+\to K^+ \bar \nu \bar \nu) & = \frac{1}{3} \left(\sum_\nu \left|1-\frac{C^{\nu}_{LQ}}{C^\nu_{SM}}\right|^2 \right) BR^{SM}(B^+\to K^+ \nu \bar\nu) \,, \\
  C^\nu_{SM} &= \frac{G_F V^*_{ts} V_{tb}}{\sqrt{2}} \frac{\alpha_{\rm em}}{2\pi \sin^2\theta_W}  
  X\left( \frac{m^2_t}{m^2_W}\right)\approx -2.81 \times 10^{-9}\nonumber
   \end{align}
 with $X(y) \approx  0.65 y^{0.575}$~\cite{Buchalla:1995vs}. Using  $|C^{LQ,\nu}_9| \sim 1$, it can be seen that the loop contribution in $C^\nu_{LQ}/C^\nu_{SM}$ is $C^{LQ,\nu}_9 C_{SM}/(4 C^\nu_{SM}) \sim 0.07$; that is, the $Z'$-mediated penguin cannot significantly contribute to $B^+ \to K^+ \nu \bar \nu$.  If we take the tree diagram as the dominant effect, to satisfy the current upper limit, the magnitude of $g_2 g^*_3$  can be in the range of $-0.096< g_2 g^*_3<0.048$.  If we ignore the small CKM matrix elements, it can be seen $g_{2}  g'^*_{3}\approx g_2 g^*_3$. Then, the contribution of the first term  in Eq.~(\ref{eq:Hbc}) to $R_{D^{(*)}}$ can be written as:
 \begin{equation}
 R_{D^{(*)}}  = \left| 1 + \delta \right|^2 R^{\rm SM}_{D^{(*)}}\,, \quad 
 \delta \approx  \frac{\sqrt{2} g_2 g^*_3 }{ 8 m^2_\Phi G_F V_{cb}}\,.
 \end{equation}
 The magnitude of $g_2 g^*_3$ can be determined as:
  \begin{equation}
  |g_2 g^*_3| \approx \frac{8 m^2_\Phi G_F |V_{cb}|}{\sqrt{2}} \left(\sqrt{\frac{R_{D^{(*)}}}{R^{\rm SM}_{D^{(*)}}}} -1 \right) \,.
    \end{equation}
 Accordingly, we obtain $|g_{2} g^*_3| \approx 0.41$ for $R_D =0.4$ and $|g_{2} g^*_3| \approx 0.25$ for $R_{D^*} =0.3$. It is clear that  the $g_2 g^*_3$ effects bounded by the $B^+ \to K^+ \nu \bar \nu$ decay cannot suffice to explain the observed $R_{D^{(*)}}$. Thus, we need to rely on the scalar- and tensor-type four-Fermi interactions shown in Eq.~(\ref{eq:Hbc}) to resolve the $R_{D^{(*)}}$ excesses. 
 
 In addition to the $B^- \to D^{(*)} \tau \bar \nu_\tau$ decays, the effective interactions in Eq.~(\ref{eq:Hbc}) can also contribute to the $B_c \to \tau \bar \nu_\tau$ decay, and the BR can be formulated as:
  \begin{align}
 BR(B_c \to \tau \bar \nu_\tau) = BR^{\rm SM}(B_c \to \tau \bar \nu_\tau)   \left| 1 + \frac{m^2_{B_c}}{m_\tau (m_b + m_c)} \epsilon_P \right|^2,
 \label{eq:BRBc}
 \end{align}
 where $BR^{\rm SM}(B_c \to \tau \bar \nu_\tau) \approx 2.1\%$,  $\epsilon_P= \sqrt{2} w_2 g^*_3/(8 G_F V_{cb} m^2_\Phi)$, and the  contribution from $g_2 g'^*_3$ has been dropped. As pointed out by the authors in~\cite{Li:2016vvp,Alonso:2016oyd,Akeroyd:2017mhr}, due to the enhancement factor $m^2_{B_c}/m_\tau (m_b +m_c) \sim 3.6$, the $B_c \to \tau \bar \nu_\tau$ decay can give a strict bound on the associated parameter. 
 Since the sign of $w_2 g^*_3$ for enhancing $R_{D^{(*)}}$ is negative, with $w_2 g^*_3\sim -0.3$, we obtain $\epsilon_P \sim -0.114$. 
In this case, we get $BR(B_c \to \tau \bar \nu_\tau) \sim 0.7 \ \%$.
It can be seen that the preferred values of $w_2 g^*_3$ will destructively interfere with the SM so that  $BR(B_c \to \tau \bar \nu_\tau) < BR^{\rm SM}(B_c \to \tau \bar \nu_\tau)$.  
Hence, the upper limit of $BR(B_c \to \tau \bar \nu_\tau) < 10\%$ obtained in~\cite{Akeroyd:2017mhr}  cannot severely bound  the LQ couplings in this model.

 The next constraint that we would like to focus on is the $\Delta F=2$ process in the neutral meson system, where the associated observable is meson mass difference $\Delta m_F$. Based on the analysis above, what we are concerned with is whether the parameters $f_2 f^*_3$ and $w_2 g^*_3$ can largely suffice to explain the $R_{K^{(*)}}$ and $R_{D^{(*)}}$ anomalies.  In order to focus on the Yukawa couplings $f_{2,3}$, $g_{3}$, and $w_2$, we can suppress the irrelevant parameters by using the scheme with $g_{1,2}, f_{1}, w_{1,3} \ll 1$. Then, we can ignore the constraints from $\Delta m_K$ and $\Delta m_{B_d}$ and only need to focus on $D-\bar D$ and $B_s -\bar B_s$ mixings. With the taken scheme, $\Delta m_{D}$ and $\Delta m_{B_s}$ can be  formulated  as:
  \begin{align}
   \Delta m^{LQ}_{D}  & \approx  \frac{ B_{D} f^2_{D} m_{D}}{3(4\pi)^2 m^2_\Phi} 
 \left|  (f'_1 f'^*_2)^2 J_2\left( \frac{m^2_L}{m^2_{\Phi}} \right)\right|\,,  \\
 \Delta m^{LQ}_{B_s}  & \approx  \frac{ B_{B_s} f^2_{B_s} m_{B_s}}{3(4\pi)^2 m^2_\Phi} 
 \left|  (f_2 f^*_3)^2 J_2\left( \frac{m^2_L}{m^2_{\Phi}} \right)\right|\,,  \\
 J_2(x) & = -\frac{1}{x-1} + \frac{x \ln x}{(x-1)^2}\,, \nonumber 
 \end{align}  
where the hadronic bag parameters $B_F$ and the meson decay constant $f_F$ are $B_{D} \approx 1.18$~\cite{Buras:2010nd}, $B_s \approx 1.28$, $f_D\approx 0.222$ GeV~\cite{PDG}, and $f_{B_s}\approx 0.231$ GeV~\cite{Lenz:2010gu}.  Due to $f_1, V_{ub}, V_{cb} \ll 1$, the Yukawa couplings $f'_{1,2}$ can be approximately expressed as $f'_1\approx f_2 V_{us}$ and $f'_2 \approx f_2$; that is $f'_1 f'^*_{2} \approx |f_2|^2 V_{us}$. It can be seen that the $D-\bar D$ mixing can directly constrain the $f_2$ parameter. Taking $\Delta m^{\rm exp}_D \approx 6.4 \times 10^{-15}$ GeV and $\Delta m_{B_s} \approx 1.17\times 10^{-11}$ GeV~\cite{PDG} as the upper bounds, the limits on $|f_2|$ and $|f_2 f^*_3|$ can be obtained as:
 \begin{align}
 |f_2| < 4.84\times 10^{-3} \left( \frac{m^2_\Phi}{J_{2}(m^2_L/m^2_\Phi)}\right)^{1/4}\,, \nonumber \\
 |f_{2} f^*_{3}| < 1.23\times 10^{-4} \left( \frac{m^2_\Phi}{J_{2}(m^2_L/m^2_\Phi)}\right)^{1/2}\,.
 \end{align}
 With $m_L/m_\Phi\sim 2$, it can be seen that $|f_2| < 0.21$ and $|f_2 f^*_3| < 0.23$. Compared to  the requirement of $f_{2} f^*_{3} \sim 0.06$ for $C^{LQ,\mu}_9 \sim -1$, the bounds from $\Delta m_{D}$ and $\Delta m_{B_s}$ are mild.

\begin{figure}[hpbt]
\includegraphics[width=90mm]{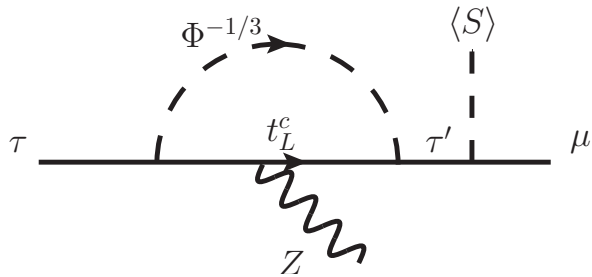}  
\caption{ Sketched Feynman diagrams for the induced $Z\tau\mu$ coupling.}
\label{fig:Ztaumu}
\end{figure}

 It has been investigated that the lepton-flavor violating (LFV) effects for explaining $R_{D^{(*)}}$ and $R_{K^{(*)}}$ excesses can be constrained by the precision measurements, such as $Z\to \tau \mu$, $\tau\to \mu (\pi, \rho)$, and $\tau\to 3 \mu$~\cite{Feruglio:2016gvd,Feruglio:2017rjo}. We examine these constraints in our model. From Eq.~(\ref{eq:Yukawa}), it can be seen that the right-handed $\tau$-lepton and left-handed muon can couple to the heavy VLL. Therefore, the coupling $Z\tau\mu$ can be generated at the tree level via the lepton Yukawa couplings. However, in order to obtain the same chiralities in both tau-lepton and muon when they couple to the $Z$-boson, one of leptons has to flip the chirality; as a result,  the tree-induced $Z\tau\mu$ coupling is suppressed by $m_{\tau,\mu}/v$. Using $y_\tau\sim  y'_\mu\sim 0.1$ and $v_S \sim 264$ GeV, the resulted BR for $Z\to  \mu \tau$ is $BR(Z\to \mu \tau) \sim 3.9 \times 10^{-13}$, which is far below the current upper limit with $BR(Z\to \mu \tau)< 1.2\times 10^{-5}$~\cite{PDG}. In addition to the tree effects, the $Z\tau\mu$ coupling can be  induced  through the loop penguin diagrams, where the main Feynman diagram is shown in Fig.~\ref{fig:Ztaumu}.  We note that since the $Z'$-boson does not couple to the quarks, the similar diagram for $Z'\tau\mu$, where the $Z'$ is emitted from the LQ, is suppressed by $m_{\tau,\mu}/v$; therefore, their effects can be neglected. Accordingly, the effective interaction for $Z\tau\mu$  can be expressed as:
 \begin{align}
 {\cal H}_{\mu \tau Z} & = \frac{g C^u_L f'_t g'^*_t}{2\cos\theta_W} \left( \frac{y'_\mu v_S}{\sqrt{2} m_L}\right) J_3\left( \frac{m^2_t}{m^2_\Phi}\right) \bar \mu \gamma_\mu P_L \tau Z^\mu  \nonumber \\
 C^u_L &=1 - \frac{4}{3} \sin^2 \theta_W\,, \quad J_3(x)  = -\frac{x}{1-x} - \frac{x \ln x}{(1-x)^2}\,,
 \end{align}
where $C^u_L$ is the Z-boson coupling to the up-type quarks. Since the induced $Z\tau\mu$ coupling is related to the up-type quark mass, we only show the top-quark contributions due to $m_{u,c}\ll m_t$. Although the $Z\tau\mu$ interaction can contribute to $Z\to \tau\mu$ and $\tau \to \mu \ell \bar \ell$,  since  the current upper limit of $Z\to \mu \tau$ is much larger than  that of $\tau\to 3\mu$ with $BR(\tau \to 3 \mu)< 2.1 \times 10^{-8}$~\cite{PDG}, we focus on the analysis of $\tau\to \mu \ell \bar \ell$, where $\ell$ can be the neutrinos and charged leptons. Thus, the $Z$-mediated BR for $\tau \to \mu \ell \bar\ell$ is given as:
\begin{align}
BR(\tau \to \mu \ell \bar\ell)& = \frac{\tau_\tau m^5_\tau G^2_F}{192 \pi^3} \left( |C^\ell_R|^2 +|C^\ell_L|^2\right) |X_{\tau\mu}|^2\,, \nonumber \\
X_{\tau \mu} & = \frac{C^u_L f'_3 g'^*_3}{2(4\pi)^2 } \left( \frac{y'_\mu v_S}{\sqrt{2} m_L}\right) J_3\left( \frac{m^2_t}{m^2_\Phi}\right)\,,
\end{align}
where $C^{\ell}_{R,L}$ are the $Z$-boson couplings to the leptons, and they are given as $C^{\nu}_R=0$, $C^\nu_L = 1$, $C^{\ell^-}_R=2\sin^2\theta_W$, and $C^{\ell^-}_L= -1+2\sin^2\theta_W$.  Using $f'\sim g'\sim 1$, $y'_\mu \sim 0.1$, and $v_S\sim 264$ GeV, we get $BR(\tau \to \mu \nu \bar \nu) \sim 9\times 10^{-10}$ and $BR(\tau \to 3 \mu)\sim 1.5 \times 10^{-10}$. It is clear that with a smaller $y'_\mu$,  the $f'_3$ and $g'_3$ parameters  can scape from the strict constraints of  the rare tau decays.

 Finally, we briefly discuss the constraints from the LQ production at the LHC~\cite{Faroughy:2016osc}. According to Eq.~(\ref{eq:Yukawa}), the LQ couplings to $\nu_\tau b$, $\tau (t,c)$ are $g_3$, $g'_3$, and $w_2$, respectively. If we assume that the scalar LQ predominantly couples to the third-generation leptons and quarks, the upper limits on the LQ mass in pair production at the LHC are as follows: using the $\nu_\tau b$  channel~\cite{Aad:2015caa}, ATLAS obtained $m_\Phi < 625$ GeV, and CMS got $m_\Phi< 634$ GeV when the $\tau t$ channel~\cite{CMS:2014gha} is used.  That is, the LQ with a mass of TeV in our case still satisfies the LHC limits, which are from the LQ-pair production~\cite{Diaz:2017lit}. In addition,  the single $\Phi^{-1/3}$ production can be produced via the $g b \to \Phi^{-1/3} \bar \nu_\tau$ and $gc \to \Phi^{-1/3} \tau^+$ processes. If we take $BR(\Phi^{-1/3} \to f)\sim 1/2$ with $f= \nu_\tau b$ and  $\tau t$, using  the values of $w_2$ and $g_3$ which can  explain $R_{D^{(*)}}$,  the single production cross section can be calculated to be around 1 fb for  $m_\Phi \sim 1$ TeV~\cite{Chen:2017hir}.   The result is below the CMS upper limit of 4.2 fb~\cite{Khachatryan:2015qda}, where the $\mu \mu j$ channel is used to search for the second-generation LQ.

\subsection{Numerical analysis for $R_{D^{(*)}}$, $C^{LQ, \mu}_9$,  muon $g-2$, and $h\to \mu \tau$ }

\begin{figure}[hptb]
\includegraphics[width=75mm]{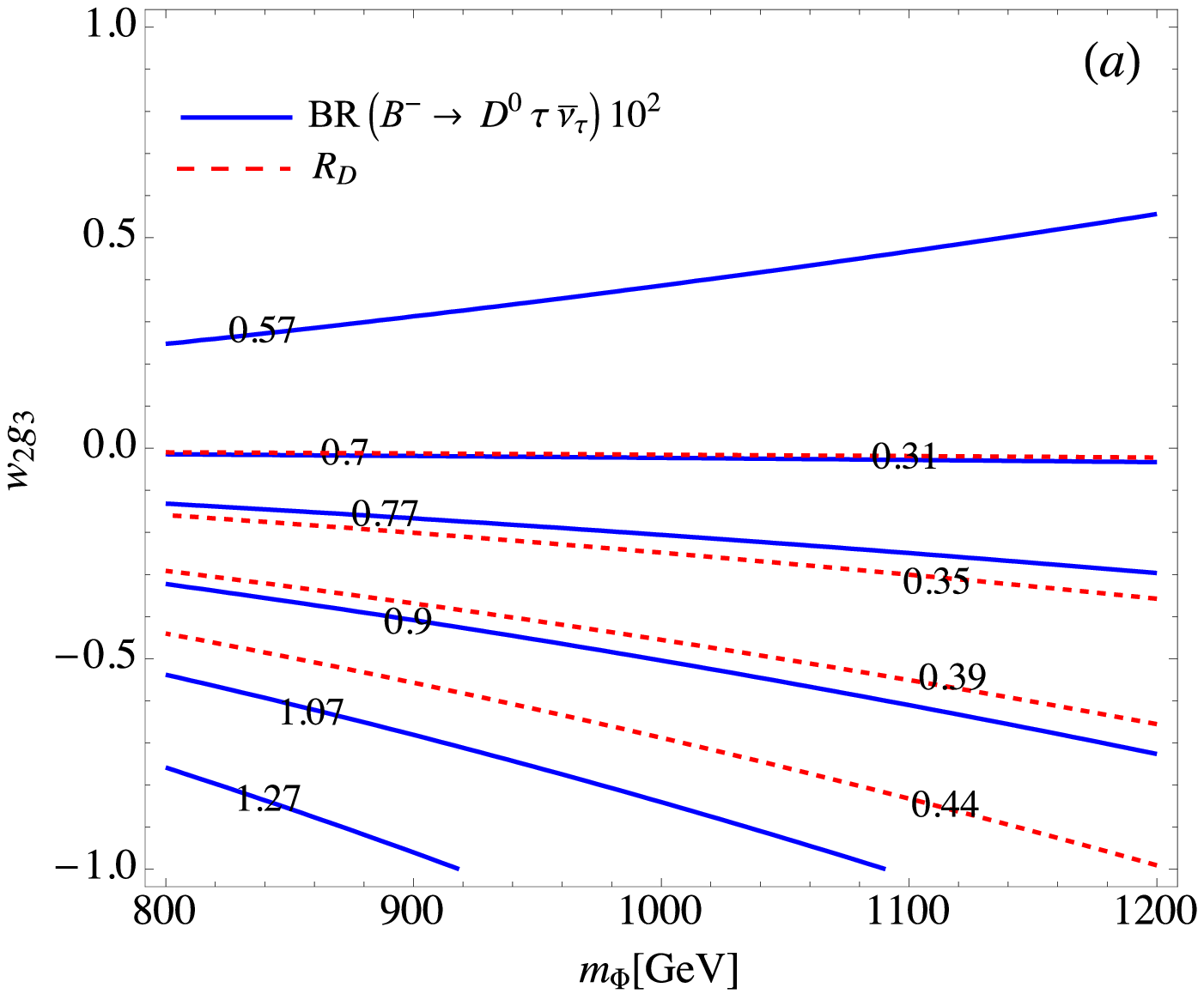}  
\includegraphics[width=75mm]{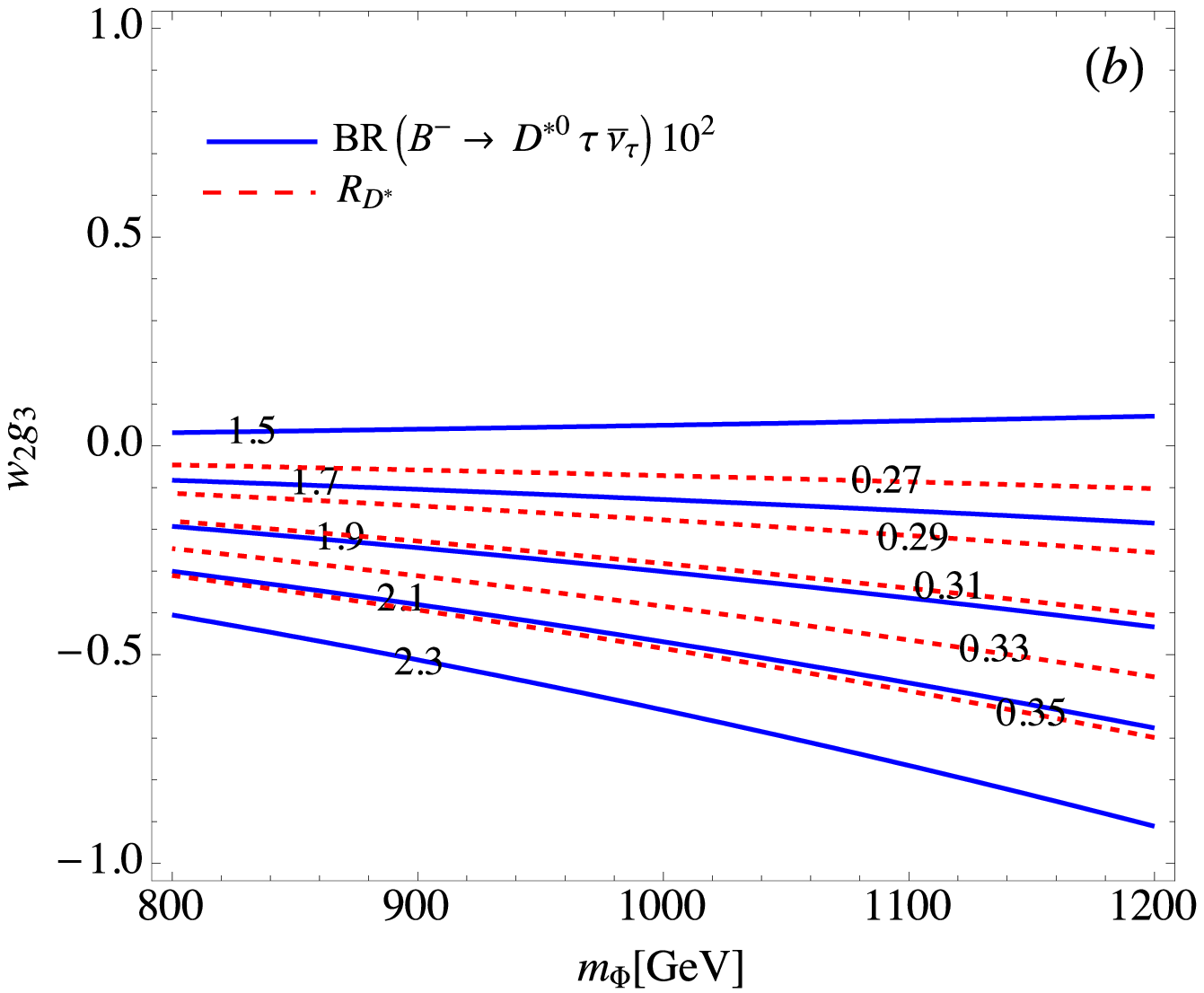} 
\caption{ Contours for (a) $BR(B^-\to D \tau \bar\nu_\tau)$ (in units of $10^{-2})$ and $R_D$ and  (b) $BR(B^-\to D^{*} \tau \bar\nu_\tau)$ (in units of $10^{-2}$) and $R_{D^*}$ as a function of $w_2 g_3$ and $m_\Phi$.}
\label{fig:fey_RD}
\end{figure}

\begin{figure}[hptb]
\includegraphics[width=85mm]{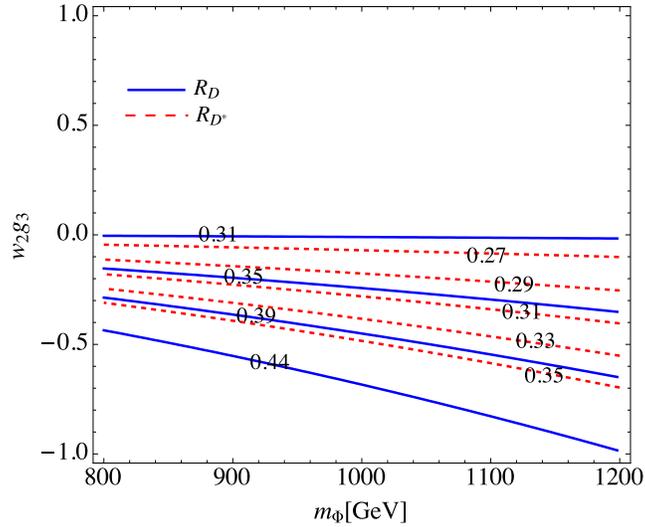}  
\caption{ Contours for $R_D$ (solid) and $R_{D^*}$ (dashed) as a function of $w_2 g_3$ and $m_\Phi$. }
\label{fig:RDRDv}
\end{figure}

To estimate the numerical results for the $B^-\to D^{(*)} \ell' \bar\nu_{\ell'}$ decays, we  use the $B\to D^{(*)}$ form factors based on the heavy quark effective theory (HQET)~\cite{Caprini:1997mu,Sakaki:2013bfa}. The BRs for $B^- \to D^{(*)} \ell' \bar\nu_{\ell'}$ in the SM are obtained as:
 \begin{align}
 BR(B^- \to [D, D^{*}] \ell \bar\nu_\ell) &\approx  [2.33\,, 5.46] \% \,,  \\
  BR(B^- \to [D, D^{*}] \tau \bar\nu_\tau) &\approx  [0.72\,, 1.39] \% \,, 
  \end{align}
where the experimental data are $BR^{\rm exp}(B^- \to [D, D^{*}] \ell \bar\nu_\ell)=[2.27 \pm 0.11,\, 5.69 \pm 0.19]\%$ and $BR^{\rm exp}(B^- \to [D,\, D^{*}] \tau \bar\nu_\tau ) = [0.77 \pm 0.25,\, 1.88 \pm 0.20]\%$~\cite{PDG}. 
   It can be seen that the $\tau \bar\nu_\tau$ measurements  are somewhat larger than the theoretical estimations. Hence,
   our  calculation  ratios $R_D$ and $R_{D^*}$ in the SM are given as:
   \begin{equation}
   R_D \approx  0.307,\, \quad R_{D^*} \approx  0.254\,.
   \end{equation}
  The obtained results are consistent with those shown in~\cite{Lattice:2015rga,Na:2015kha,Fajfer:2012vx,Bigi:2017jbd, Bernlochner:2017jka}. 
 To understand the  influence of scalar LQ on the $b\to c \tau \bar\nu_\tau$ decay, we show the contours for $BR(B^-\to [D,D^*] \tau \bar\nu_\tau)$ and $R_{D,D^*}$ as a function of $w_2 g_3$ and $m_\Phi$ in Fig.~\ref{fig:fey_RD} (a) and (b), respectively, where due to the $B^+\to K^+ \nu \bar\nu$ constraint, we have ignored the $g_2 g'^*_3$ contributions, and the renormalization group (RG) running effects from LQ scale to $m_b$ scale have been included~\cite{Sakaki:2013bfa}. From the results,  when  $R_{D}$ and $R_{D^*}$ are enhanced by the singlet scalar LQ, the $BR(B^-\to D^{(*)} \tau \bar\nu_\tau)$ can be consistent with the current data within $2\sigma$ errors. In addition, we also put  $R_{D}$ (solid) and $R_{D^*}$ (dashed) together as a function of $w_2 g_3$ and $m_\Phi$ in Fig.~\ref{fig:RDRDv}. From the plot, it can be clearly seen that the LQ contributions can simultaneously explain the $R_D$ and $R_{D^*}$ excesses in the same parameter region. 
 
\begin{figure}[hptb]
\includegraphics[width=75mm]{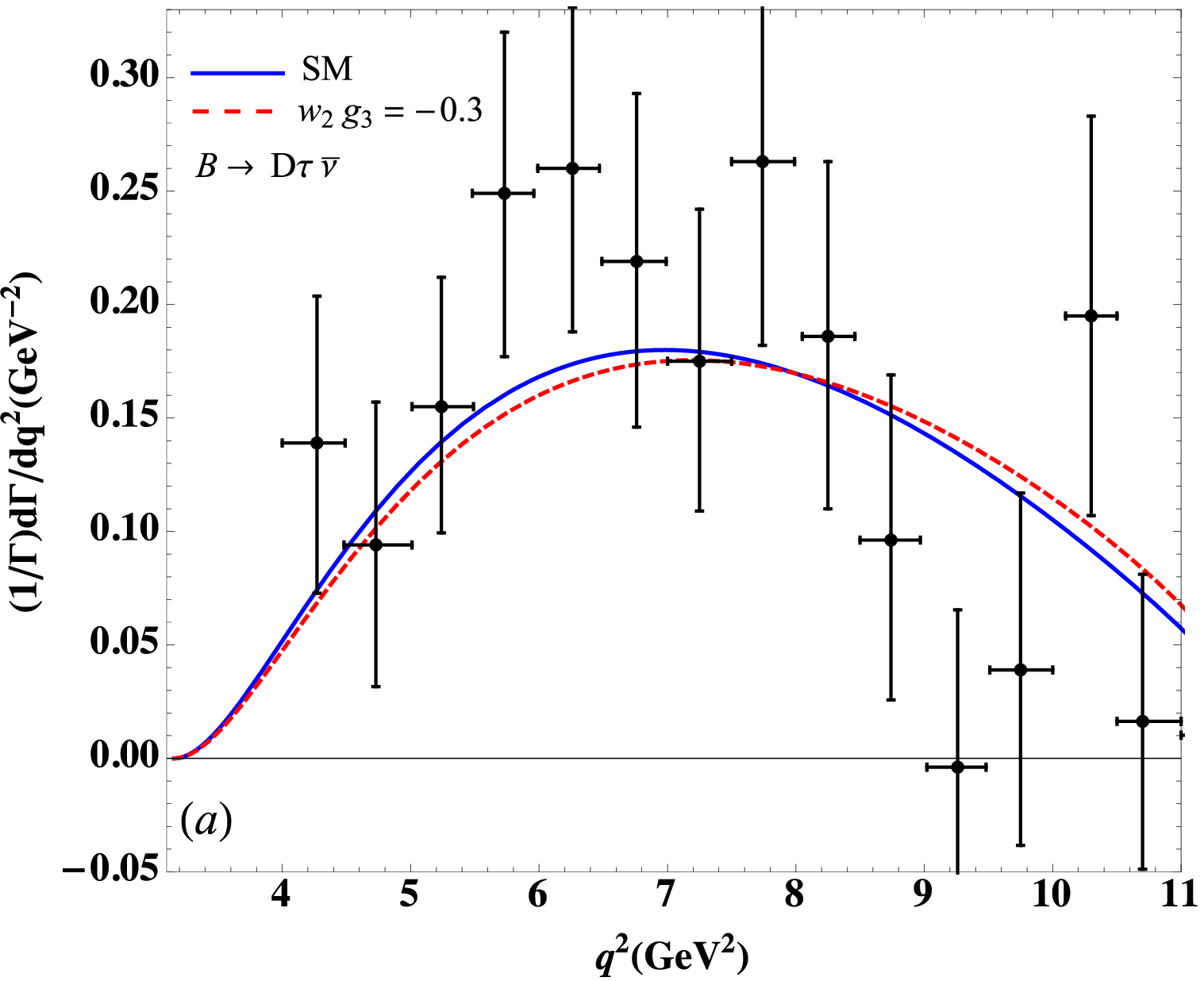}  
\includegraphics[width=75mm]{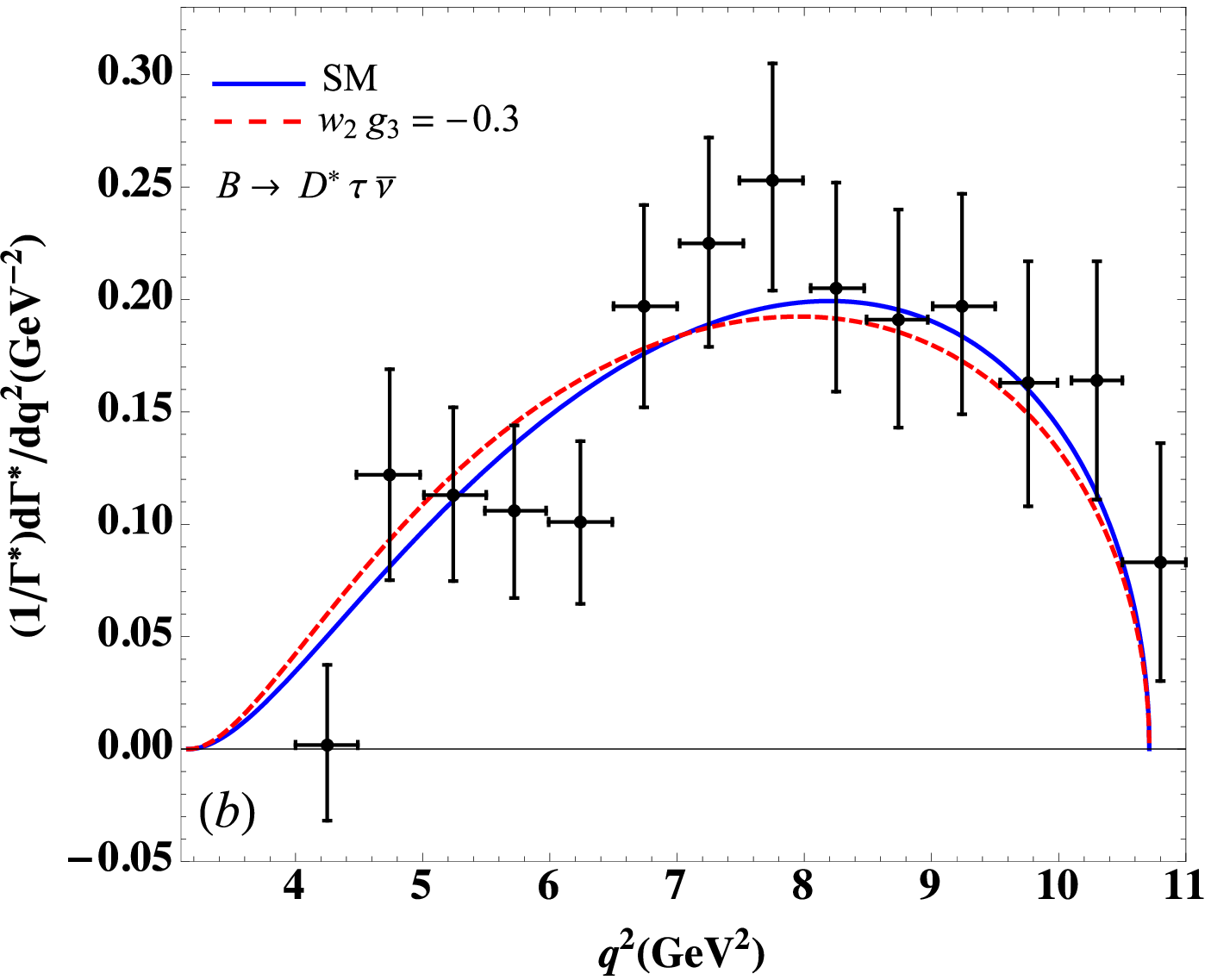} 
\caption{  $(1/\Gamma)d\Gamma/dq^2$ (in units of GeV$^{-2}$) with respect to $q^2$ for (a) $B\to D \tau \bar \nu_\tau$ and (b) $B\to D^* \tau \bar\nu_\tau$, where the solid, dashed, and dot-dashed  lines denote the results of the SM and LQ with $w_2 g^*_3=-0.2, -0.3$, respectively.  The BarBar data  are quoted from~\cite{Bardhan:2016uhr}.}
\label{fig:Diff}
\end{figure}

In addition to the $R_{D^{(*)}}$ observables,  BaBar also reported  $q^2$ distributions of the detected events for $B\to (D,D^*) \tau \nu$ in \cite{Lees:2013uzd}.  To understand the LQ influence, we show  $(1/\Gamma)d\Gamma/dq^2$ as a function of $q^2$ for $B\to D \tau \bar\nu_\tau$ in Fig.~\ref{fig:Diff}(a) and for $B\to D^* \tau \bar\nu_\tau$ in Fig.~\ref{fig:Diff}(b), where the solid, dashed, and dot-dashed lines denote the results of the SM  and LQ with $w_2 g^*_3 = -0.2, -0.3$, respectively. The experimental data  are obtained from~\cite{Bardhan:2016uhr}. It can be seen that the LQ curves are  slightly different from the SM results. That is, the $q^2$ distribution of the differential decay rate may not be a good candidate for testing the new physics.  Belle recently measures the tau-lepton polarization, defined as $P_\tau=(\Gamma^{h=+}-\Gamma^{h=-1})/(\Gamma^{h=+}+\Gamma^{h=-1})$, in $B\to D^* \tau \bar \nu_\tau$, and the result is $P_\tau(D^*)=-0.38\pm0.51^{+0.21}_{-0.16}$~\cite{Hirose:2016wfn}, where the SM prediction is  $P^{\rm SM}_{\tau} \approx -0.497$~\cite{Tanaka:2012nw,Bardhan:2016uhr}.  According to the formulas in~\cite{Bardhan:2016uhr}, we find that the LQ contributions to tau polarization in $B\to D^* \tau \bar\nu_\tau$ are $P_\tau(D^*)\approx -0.488$ for $w_2 g^*_3=-0.2$ and $P_\tau(D^*)\approx -0.479$ for $w_2 g^*_3=-0.3$. Clearly, $P_\tau(D^*)$ is not sensitive to the LQ effects in our model. In addition, we also calculate the tau polarization in $B\to D \tau \bar\nu_\tau$ as $P_\tau(D)\approx 0.401$ for $w_2 g^*_3=-0.2$ and $P_\tau(D)\approx 0.434$ for $w_2 g^*_3=-0.3$, where the SM result is $P^{\rm SM}_\tau(D)\approx 0.324$. In our model, the deviation of $P_\tau(D)$ from the SM can be $\sim 30\%$.

\begin{figure}[hptb]
\includegraphics[width=85mm]{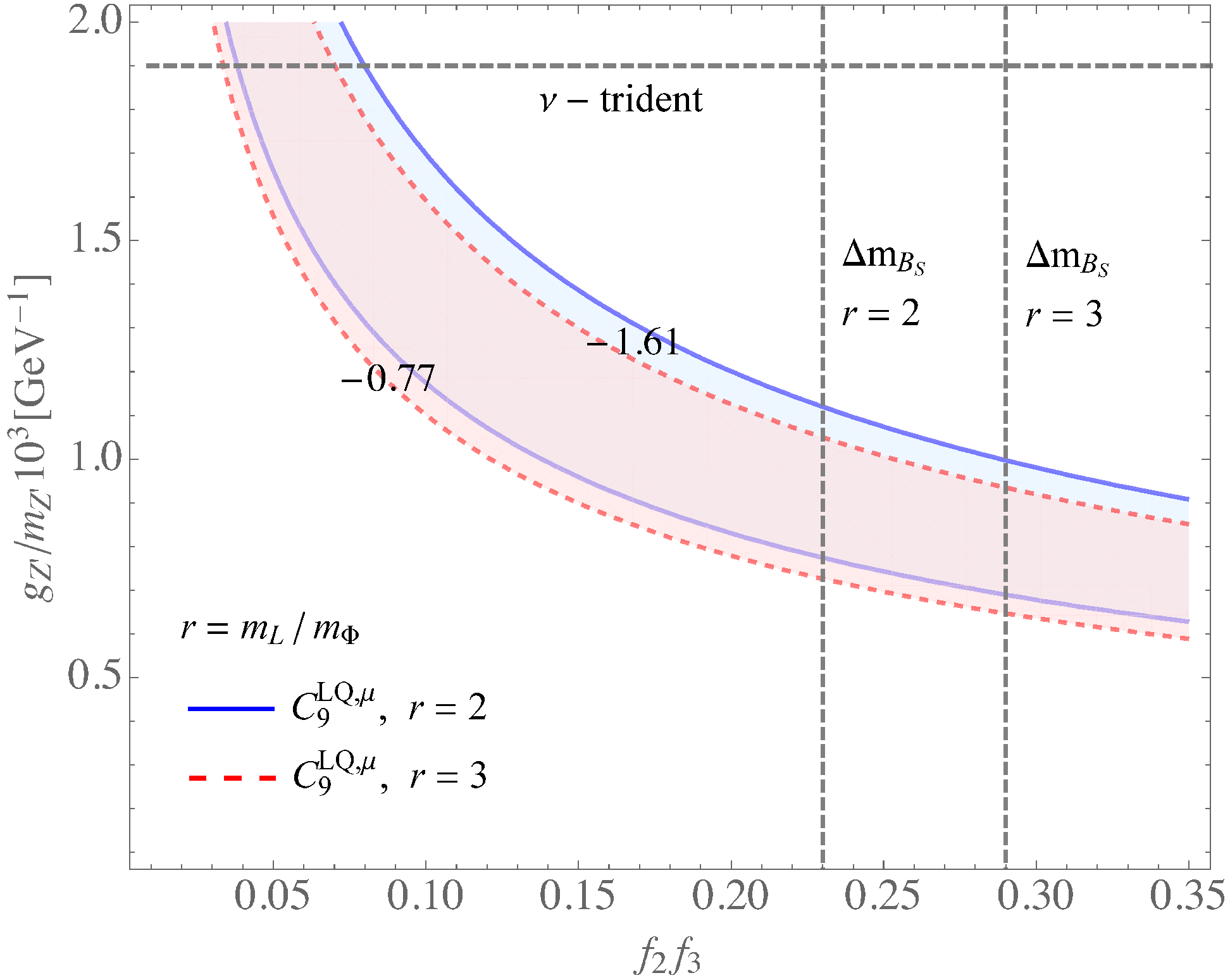}  
\caption{ Contours for $C^{LQ,\mu}_9$ as a function of $f_2 f_3$ and $g_{Z'}/m_{Z'}$ (in units of $10^{-3}$), where blue and red region respectively  denote $m_{L}/m_{\Phi}=2$ (solid) and $3$ (dashed). The bounds from the neutrino trident production and $\Delta m_{B_s}$ with $m_L/m_{\Phi}=2, 3$ are also given. }
\label{fig:fey_C9}
\end{figure}

To analyze the $R_K$ and $R_{K^*}$ issues, we take the results obtained from a global fitting to the experimental data ~\cite{Altmannshofer:2017fio}, where the required Wilson coefficient $C^{NP}_9$  from new physics  used to explain the $b\to s \mu^+ \mu^-$ anomalies can be in the region of $C^{NP}_{9}=[-1.61,-0.77]$. According to Eq.~(\ref{eq:CZp}), we then show $C^{LQ,\mu}_9$ as a function of $f_2 f_3$ and $g_{Z'}/m_{Z'}$ in Fig.~\ref{fig:fey_C9}, where the solid and dashed lines denote  $m_L/m_\Phi=2$ and $m_L/m_\Phi=3$, respectively; the constraints from the neutrino trident production and $\Delta m_{B_s}$ are included, and  the shown range for the Wilson coefficient is taken as $C^{LQ,\mu}_9\subset [-1.61,-0.77]$.  From the plot, it can be seen that the allowed parameter spaces are still wide. 

After showing the contributions to $R_{D^{(*)}}$ and $C^{LQ,\mu}_9$ in the model,  in the remainder of this paper, we discuss some interesting implications on the muon $g-2$ and the Higgs  LFV  $h\to \mu \tau$ decay. Since we concentrate on the case with $m_{Z'}\gg m_\mu$, the $Z'$-mediated muon $g-2$ can be simplified as:
  \begin{equation}
  \Delta a^{Z'}_\mu \approx \frac{g^2_{Z'} m^2_\mu}{12 m^2_{Z'}} < 3.4 \times 10^{-10}\,, \label{eq:muong2}
  \end{equation}
 where the upper limit is from the neutrino trident production constraint~\cite{Kile:2014jea,Altmannshofer:2014pba}. Although the resulted muon $g-2$ is smaller than the current experimental value of $\Delta a_\mu =(28.7\pm 8.0) \times 10^{-10}$~\cite{PDG},  a factor of around 5 improved measurements will be performed in the E989 experiment at Fermilab~\cite{Grange:2015fou} and the E34 experiment at J-PARC~\cite{Otani:2015jra}. The result in Eq.~(\ref{eq:muong2}) falls within $3\sigma$ accuracy of the new muon $g-2$ measurements and can still be tested.

 In addition to the explanation of the $R_{D^{(*)}}$ and $R_{K^{(*)}}$ anomalies, from Eq.~(\ref{eq:Yukawa}), the Higgs lepton-flvaor changing $h\to  \mu \tau$ decay can be induced at the tree level in this simple model, and the associated BR can be expressed as: 
 \begin{equation}
 BR(h\to \mu\tau) \approx  \frac{m_h v^2_S |y_\tau y'_\mu|^2}{32 \pi m^2_L \Gamma_h}\,,
 \end{equation}
 where $\Gamma_h\approx 4.21$ MeV is the Higgs width. 
 From the limit of the neutrino trident production and $m_{Z'}=2 g_{Z'} v_S$, it is known $v_S \geq 263$ GeV. Thus, with $v_S\sim 264$ GeV and $y_\tau\sim y'_\mu\sim 0.1$, the BR for $h\to \mu \tau$ can be $BR(h\to \mu \tau) \sim 0.20\%$, which is close to the CMS upper bound of  $BR^{\rm exp}(h\to \mu \tau) < 0.25\%$~\cite{CMS:2017onh}.  Intriguingly, the tree-induced coupling $h\mu \tau$ can generate the radiative LFV $\tau \to \mu \gamma$ process via Higgs-mediated one-loop and two-loop Feynman diagrams~\cite{Falkowski:2013jya,Dorsner:2015mja}.  Since the one-loop effects are suppressed by the factor $m_\ell/v$, which is from the SM coupling $h\ell\ell$, the loop-induced BR for $\tau\to \mu \gamma$ is dominated by the two-loop effects.  With above values of parameters and the results in~\cite{Dorsner:2015mja}, we obtain $BR(\tau \to \mu \gamma) \sim 3.2 \times 10^{-10}$, and the result is well below the current experimental upper limit with $BR(\tau \to \mu \gamma)< 4.4 \times 10^{-8}$~\cite{PDG}.  We note that $\tau\to \mu \gamma$ can be also produced through top-quark and LQ loop, where the  related couplings are $w_3$, $f'_3$, and $y'_\mu$, since we have taken $w_3 \ll 1$,  such loop contribution could be taken to be small.

 \section{Summary}

We studied the $U(1)_{L_\mu - L_\tau}$ extension of the SM to resolve the $R_{D^{(*)}}$ and $R_{K^{(*)}}$ anomalies.   In order to achieve this purpose, we introduce one vector-like doublet lepton, one scalar leptoquark, and one singlet scalar, in which they all carry $U(1)_{L_\mu-L_\tau}$ charges. As a result, the $b\to s \mu^+ \mu^-$ process can arise from the $Z'$-penguin diagram via the leptoquark loop, whereas the $b\to c \tau \bar\nu_\tau$ decay can be induced from the same leptoquark.  When considering the constraints from the flavor physics, such as the  $b\to s \gamma$, $B^+ \to K^+ \nu \bar\nu$, $B_c\to \tau \bar \nu_\tau$, $\Delta F=2$, and $\tau \to \mu \ell \bar \ell$ processes, it is found that $R_{D}$ and $R_{D^*}$ can simultaneously fit the data in the same parameter space, and the $Z'$-penguin induced Wilson coefficient can be $C^{LQ,\mu}_9=[-1.61,-0.77]$, for which the result is from the $\chi^2$ analysis and can be used to explain the $b\to s \mu^+ \mu^-$ anomalies. In this model, due to the vector $Z'$ coupling, the Wilson coefficient $C^{LQ,\mu}_{10}$ automatically vanishes; therefore, the rare $B_s \to \mu^+ \mu^-$ process cannot give a strict bound on the parameters. The BR for the Higgs lepton-flavor violating $h\to \mu \tau$ decay can be as large as the current experimental upper limit. The $Z'$-mediated muon $g-2$ can reach the $3\times 10^{-10}$ level, which can be tested in  future new muon $g-2$ experiments.  In addition, we find that the tau polarization in $B\to D \tau \bar\nu_\tau$ is more sensitive to  the LQ effects and can have a deviation of $30\%$  in our model.

\section*{Acknowledgments}

This work was partially supported by the Ministry of Science and Technology of Taiwan
R.O.C.,  under grant MOST-103-2112-M-006-004-MY3 (CHC).

\end{document}